\UseRawInputEncoding

\documentclass[aps,prl,twocolumn,amsmath,amssymb,nofootinbib,superscriptaddress,floatfix]{revtex4-1}

\usepackage{amsmath}
\usepackage{amssymb}
\usepackage{amsthm}
\usepackage{mathtools}
\usepackage{mathdots}
\usepackage{amsfonts}
\usepackage{bbm}
\usepackage{xr}
\usepackage{bbold}
\usepackage{newpxtext}
\usepackage{epsfig,color,dsfont,upgreek,physics}
\usepackage{mathrsfs}
\usepackage{multirow}

\usepackage{graphicx}
\usepackage{dcolumn} 
\usepackage{bm} 
\usepackage{float} 
\usepackage[driverfallback=dvipdfm]{hyperref} 
\usepackage{longtable}
\usepackage{ulem}   
\allowdisplaybreaks

\newcommand{\bs}[1]{{\boldsymbol{#1}}}

\begin{document}

\title{
Emergence of Chern Supermetal and Pair-Density Wave through Higher-Order Van Hove Singularities in the Haldane-Hubbard Model
}

\author{Pedro Castro}
\affiliation
{
Department  of  Physics,  Emory  University,  400 Dowman Drive, Atlanta,  GA  30322,  USA
}

\author{Daniel Shaffer}
\affiliation
{
Department  of  Physics,  Emory  University,  400 Dowman Drive, Atlanta,  GA  30322,  USA
}

\author{Yi-Ming Wu}
\affiliation
{
Stanford Institute for Theoretical Physics, Stanford University, Stanford, California 94305, USA  
}

\author{Luiz H. Santos}
\affiliation
{
Department  of  Physics,  Emory  University,  400 Dowman Drive, Atlanta,  GA  30322,  USA
}

\begin{abstract}
While advances in electronic band theory have brought to light new topological systems, understanding the interplay of band topology and electronic interactions remains a frontier question. In this work, we predict new interacting electronic orders emerging near higher-order Van Hove singularities present in the Chern bands of the Haldane model. We classify the nature of such singularities and employ unbiased renormalization group methods that unveil a complex landscape of electronic orders, which include ferromagnetism, density-waves and superconductivity. Importantly, we show that repulsive interactions can stabilize long-sought pair-density-wave state and an exotic Chern supermetal, which is a new class of non-Fermi liquid with anomalous quantum Hall response. This framework opens a new path to explore unconventional electronic phases in two-dimensional chiral bands through the interplay of band topology and higher-order Van Hove singularities.

\end{abstract}

\date{\today}

\maketitle

\noindent

\noindent
\textit{Introduction--} 
The Haldane model \cite{Haldane1988} provides a minimal description of Chern bands displaying quantized Hall conductance albeit in zero magnetic field, which are realized when Dirac fermions in graphene bands acquire a time-reversal broken, inversion symmetric mass at the two valleys $\pm \bs{K}$ of the Brillouin zone (BZ). Besides the realization of Haldane Chern bands in ultracold fermions \cite{jotzu2014experimental}, new correlation-driven Chern bands recently identified \cite{Spanton2018,sharpe2019emergent,serlin2020intrinsic,Saito2021,xie2021fractional,nuckolls2020strongly,wu_chern_2021,das2021symmetry,choi2021correlation,park2021flavour,stepanov2021competing,pierce2021unconventional,yu2021correlated} in moir\'e materials open new paths to explore interaction effects in topological bands. 
The role of interactions in Haldane Chern bands has received substantial attention. In particular, in the strong coupling regime obtained when the interaction scale $U$ is much stronger than the bandwidth $W$ of a Chern band that is separated from other bands by an energy gap $\Delta$ such that $\Delta \gg U \gg W$, electronic correlations can stabilize rich fractionalized phases \cite{Neupert-2011,Sheng-2011,Tang-2011,Sun-2011}. Furthermore, correlated phases of the repulsive Haldane-Hubbard model at commensurate $1/2$ and $1/4$ fillings have been investigated using numerical methods and mean-field studies \cite{MaiFeldmanPhillips23,He11,Zheng15, wu2016quantum,ArunSohalParamekanti16,VanhalaTorma16, ImriskaTroyer16, tupitsyn2019phase,Yi21}.

On the other hand, defying expectations that the weak coupling regime $(U \ll W)$ in Chern bands leads to a stable Fermi liquid (FL) fixed point, recent analysis \cite{ShafferUnconventional2022,shaffer2022triplet} has uncovered new FL instabilities in partially filled Hofstadter bands \cite{Hofstadter76} (generally supporting non-zero Chern number \cite{TKNN}) due to the interplay of repulsive interactions and logarithmic Van Hove singularities (VHS). In such fractal bands, the magnet flux per unit cell acts a control knob changing the landscape of VHSs and opening new interaction channels. A pressing question then arises: Can VHS catalyze FL instabilities in Haldane Chern bands in the absence of a magnetic field? Notably, while the relation between VHS and electronic correlations has been greatly emphasized in modern 2D materials, from moir\'e \cite{li2010observation,kerelsky2019maximized,wu2021chern} to kagom\'e metals \cite{kang2022twofold,neupert2022charge}, little is known about their influence in Chern bands.

\begin{figure}
    \centering
    \includegraphics[width=8.5cm]{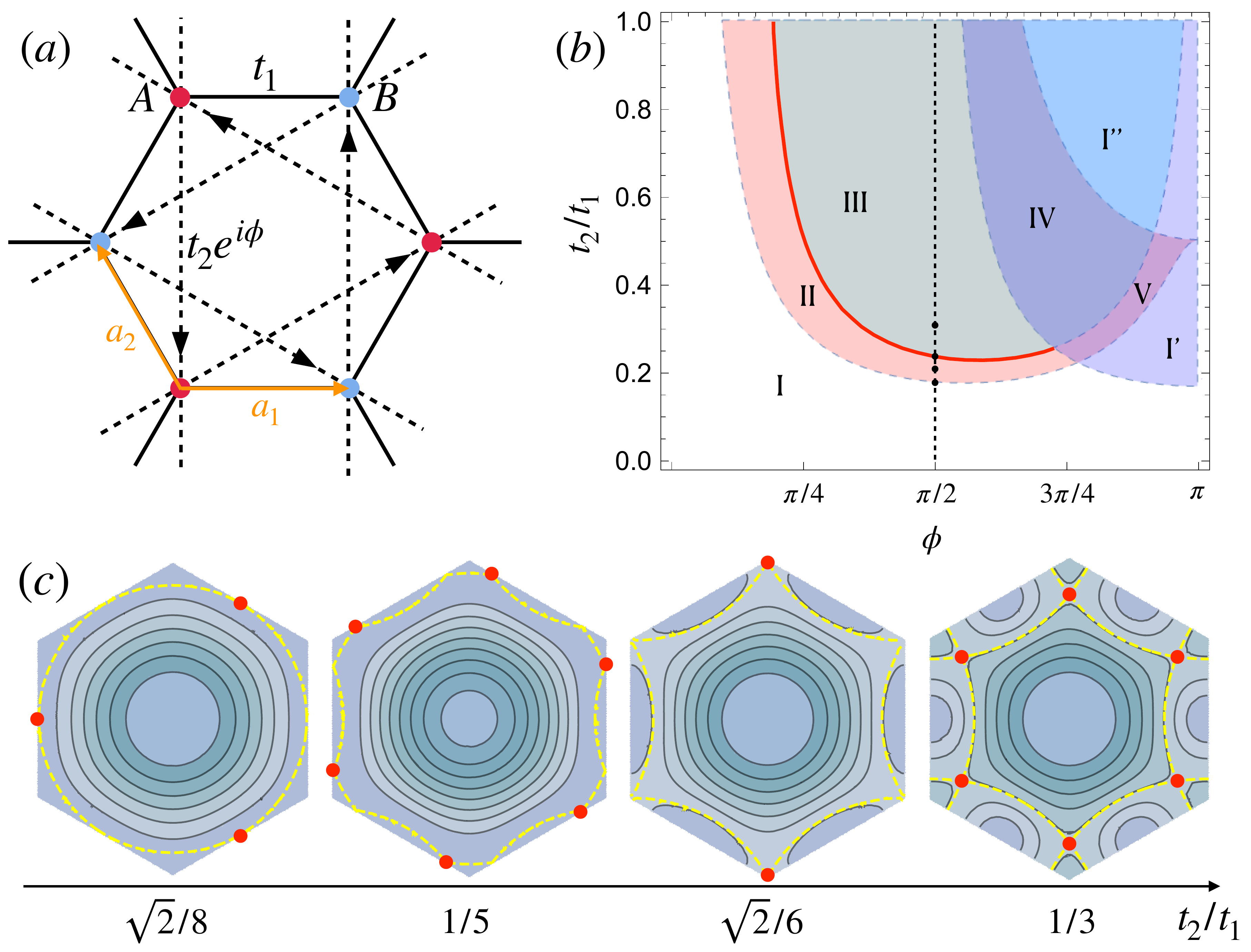}
    \caption{(a) Lattice model. (b) Landscape of VHSs. The boundary between II and III highlighted in red is where two higher order VHSs located at $\pm\bm{K}$ emerge. (c) Energy contours show the evolution of VHSs for a fixed $\phi=\pi/2$ but various $t$, which are marked as the black dots in (b). The yellow dashed curves are the Fermi surfaces at Van Hove filling.}
    \label{fig:intro}
\end{figure}

In this Letter, we unveil a new scheme to investigate correlation effects in Chern bands, focusing on the Haldane-Hubbard model as a paradigmatic system to address the confluence of band topology and electronic correlations. Departing from previous studies
\cite{MaiFeldmanPhillips23,He11,Zheng15, wu2016quantum,ArunSohalParamekanti16,VanhalaTorma16, ImriskaTroyer16, tupitsyn2019phase,Yi21}, we analyze a new regime characterized by incommensurate fillings reached when the Fermi energy lies near VHS in the Haldane Chern bands. The diverging density of states (DOS) near localized pockets in the BZ allows a treatment of interactions at weak coupling regime using unbiased RG methods.
The main results of this work are:
\\
\noindent
(1) \textit{A novel classification of VHS in Haldane Chern bands:}
We analytically identify new
logarithmic \cite{vanHove1953theoccurrence} and higher-order VHS (HOVHS) \cite{shtyk2017electrons,yuan2019magic} of Haldane Chern bands, beyond the conventional VHSs in graphene \cite{neto2009electronic}. These new saddle points are controlled by the second neighbor hopping amplitude $t_2$ and the phase $\phi$ (see Fig. \ref{fig:intro}.), which break time reversal symmetry \textit{while preserving spatial inversion.} Notably, we identify a pair of HOVHS at $\pm \bs{K}$ occurring throughout the boundary of regions II and III of Fig. \ref{fig:intro}. Such HOVHSs yield diverging DOS $ D(\varepsilon) \sim |\varepsilon|^{-1/3}$, promoting strongly enhanced low temperature susceptibilities in all particle-particle and particle-hole channels, which, consequently, open a new path to explore competing electronic orders in Chern bands via these HOVHS. 
\\
\noindent
(2) \textit{Novel FL instabilities through competing orders near HOVHS:}
We employ perturbative RG \cite{shankar_renormalization-group_1994,maiti_superconductivity_2013} to study FL instabilities in the vicinity of such pair of HOVHS related by inversion symmetry. 
While this HOVHS 2-patch RG was analyzed in bilayer graphene \cite{shtyk2017electrons} and moir\'e systems \cite{HsuDasSarma21,wu2023pair},
the situation in Haldane Chern bands is distinct in that \textit{band topology non-trivially constraints the RG flows.} 
This occurs because the topological winding ($\pm 1$) of Haldane Chern bands is associated with electronic wavefunctions on opposite $\pm \bs{K}$ valleys which have support on opposite sublattices. This entails a form of sublattice interference (SI), which suppresses certain interaction 
channels,
profoundly modifying the RG flows and the resulting orders. While this SI was first noticed in the kagome lattice \cite{kiesel2012sublattice, wu2022sublattice}, to our knowledge, the existence of this important effect and its connection to competing orders in Haldane Chern bands has not received earlier consideration, and is one of the central results of this work.

\begin{figure}
    \centering
    \includegraphics[width=\columnwidth]{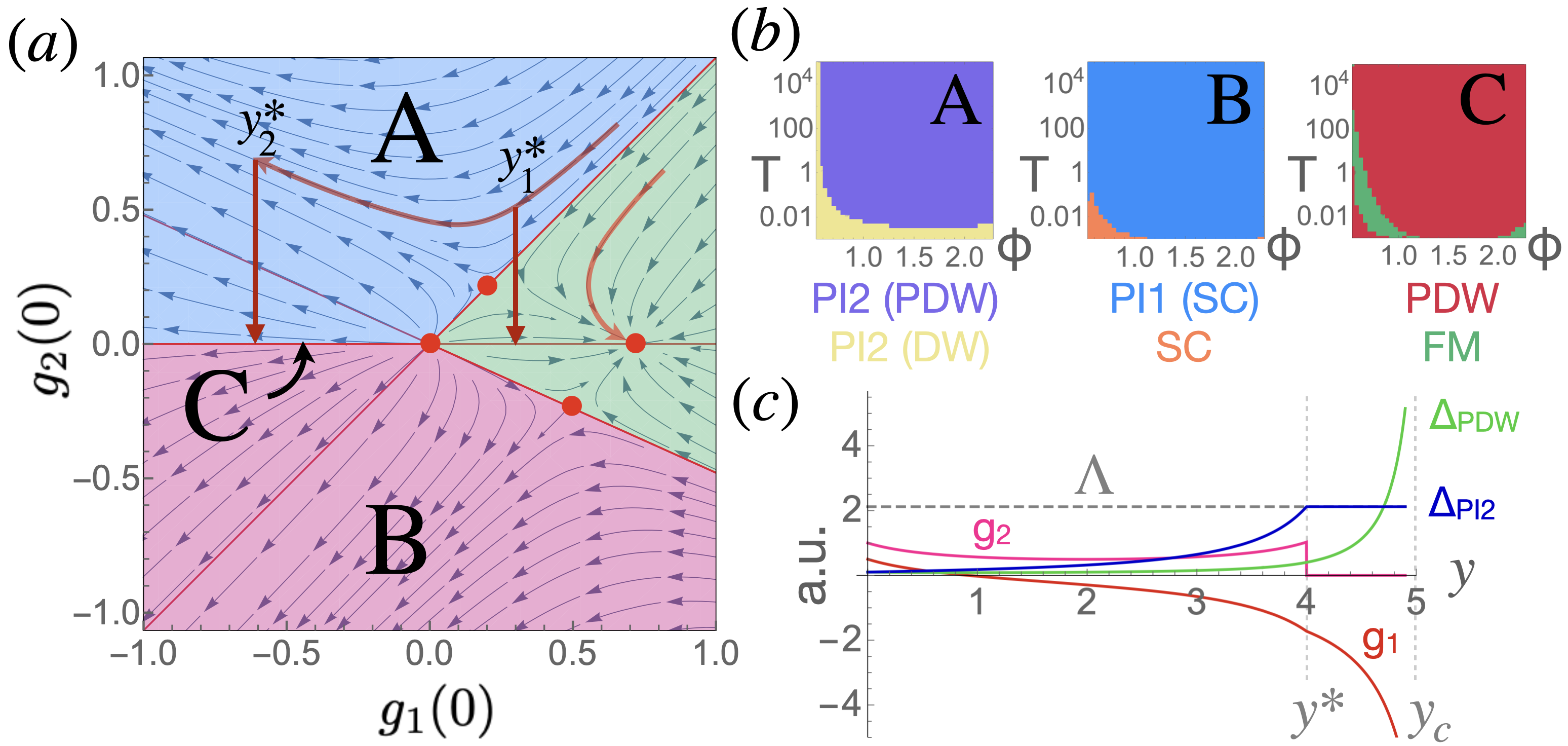}
    \caption{(a) \( g_1(0)=U,\, g_2(0)=V\) phase diagram and RG flow trajectories. Fixed points are shown by red dots, fixed trajectories are shown by red lines that also form the phase boundaries. All trajectories in the green region flow to the supermetal fixed point, while all trajectories in the blue region A with \(g_2>0\) (red region B with \(g_2<0\)) flow to some symmetry-broken strong coupling fixed point. The corresponding phases depend on temperature \(T\) and \(\phi\), as shown in (b): the PI2 instability is always leading in A with a subleading PDW (purple) or C/SDW (yellow, labeled DW) instability; either a leading PI1 with subleading \(s/t\)SC or a leading \(s/t\)SC occur in B. The fixed trajectory along \(g_2=0\) is unstable but can become stabilized at \(y=y^*\) due to a staggered mass generated by the PI2 instability becoming comparable to the RG cutoff \(\Lambda\), as shown in (c), which also shows the RG flows of \(g_1\) and \(g_2\) starting in region A with repulsive bare interactions. Depending on whether \(g_1\) changes sign or not at \(y^*\) (\(y^*_2\) and \(y^*_1\) labeled in (a)), either an instability in region C (PDW or FM) or a supermetal phase is realized respectively.}
    \label{fig:phasediagramflows}
\end{figure}

Incorporating this novel SI into RG analysis brings forth a rich phase diagram, shown in Fig.\ref{fig:phasediagramflows}, containing a host of ordered phases including ferromagnetism (FM), density-wave (DW), superconductivity (SC) and pair-density-wave (PDW). Remarkably, we also identify a mechanism whereby repulsive interactions can stabilize an exotic interacting supermetal fixed point \cite{IsobeFu19}, which arises here as a non-Fermi liquid with quantum anomalous Hall response \cite{nagaosa2010anomalous}. We coin this new phase a \textit{Chern supermetal}.
While the supermetal in \cite{IsobeFu19} relies on an isolated HOVHS (see also \cite{AksoyChamon23}), we demonstrate that SI and spontaneous generation of a staggered mass under the RG flow in the Haldane Chern bands is pivotal in suppressing interaction channels between the two HOVHS that would normally drive the system towards an ordered phase \cite{shtyk2017electrons,classen_competing_2020}. The same mechanism also allows for a pair-density-wave (PDW) state \cite{agterberg2020physics} with momentum $\pm \bs{K}$ to emerge from purely repulsive interactions. Our results thus constitute a new paradigm to explore electronic in correlations topological Chern bands via HOVHS.

\noindent \textit{Model}. We investigate the spin-degenerate single particle Hamiltonian \cite{Haldane1988}
\begin{equation}
\label{eq: Haldane model}
H_0 = \sum_{\sigma = \uparrow,\downarrow} \sum_{\bs{k} \in \textrm{BZ}}
c^{\dagger}_{\bs{k}\sigma} \mathcal{H}_{\bs{k}} c^{}_{\bs{k}\sigma}
\,,
\quad
\mathcal{H}_{\bs{k}} = B_{0,\bs{k}}\tau_0 + \bs{B}_{\bs{k}}\cdot\bs{\tau}
\,,
\end{equation}
where $\textrm{BZ}$ is the first Brillouin zone, $c_{\bs{k}\sigma} = (c_{A,\bs{k},\sigma}, c_{B, \bs{k}, \sigma})^{T}$ is the fermionic operator on sublattices $A$ and $B$;  $\tau_{\mu}$ ($\mu=0,1,2,3$) represents the $2\times 2$ identity and three Pauli matrices acting on the sublattice degrees of freedom, and
\begin{align}
&
B_{0,\bs{k}} = 2t_2\cos\phi\sum_{i=1}^3\cos\bs{k}\cdot\bm{b}_i\,, \nonumber \\
&\,
\bs{B}_{\bs{k}}=
\sum_{i = 1}^{3}
\begin{pmatrix}
t_1
\cos\bs{k}\cdot\bm{a}_i
\\
-t_1
\sin\bs{k}\cdot\bm{a}_i
\\
2t_2\sin\phi
\sin\bs{k}\cdot\bm{b}_i
\end{pmatrix}
\label{PauliCoefficients}
\,,
\end{align}
where
$\bm{a}_1=a(1,0), \ \bm{a}_2=\tfrac{a}{2}(-1,\sqrt{3}), \ \bm{a}_3=\tfrac{a}{2}(-1,-\sqrt{3})$ are vectors connecting A to nearest neighbor B sites and
$\bm{b}_1=\bs{a}_2 - \bs{a}_3, \bm{b}_2= \bs{a}_3 - \bs{a}_1, \bm{b}_3= \bs{a}_1 - \bs{a}_2$. $a$ is the lattice constant that we henceforth set to one and we define $t = t_2/t_1$.
The single particle Hamiltonian \ref{eq: Haldane model} is diagonalized $H_0 = \sum_{\sigma}\sum_{n = \pm}\,\varepsilon_{n,\bs{k}}\psi^{\dagger}_{n,\sigma,\bs{k}}\psi^{}_{n,\sigma,\bs{k}}$ by the unitary transformation $c_{s,\bs{k},\sigma} = \sum_{n = \pm} u_{s n}(\bs{k})\psi_{n,\bs{k},\sigma}$, ($s= A,B$) leading to spin degenerate energy bands
$
\varepsilon_{\pm, \bs{k}}
=
B_{0,\bs{k}} \pm |\bs{B}_{\bs{k}}|
\,.
$
In order to classify the critical points $\bs{\nabla}_{\bs{k}} \varepsilon_{\pm, \bs{k}} = 0$, we analyse the determinant of the Hessian
$
 \mathbb{H}_{n,\bs{k}}   
 =
 \textrm{det}\Big( \frac{\partial^2 \varepsilon_{n,\bs{k}}}{\partial k_{i} \partial k_{j}} \Big)
 \,,
$
which yields the local minima/maxima ($\mathbb{H}_{n,\bs{k}} > 0$), conventional saddle points ($\mathbb{H}_{n,\bs{k}} < 0$) and higher order saddles ($\mathbb{H}_{n,\bs{k}} = 0$). 

Henceforth we focus on the upper band $\varepsilon_{+}$ with analogous considerations holding for $\varepsilon_{-}$. The model at $t=0$ reduces to graphene \cite{neto2009electronic} whose critical points are $\bm{\Gamma}=(0,0)$, $\pm \bm{K}= \pm \left(\frac{2\pi}{3},\frac{2\pi}{3\sqrt{3}}\right)$, $\bm{M}_1=\left(\frac{2\pi}{3},0\right)$, $\bm{M}_2=\left(\frac{\pi}{3},\frac{\pi}{\sqrt{3}}\right)$ and $\bm{M}_3=\left(-\frac{\pi}{3},\frac{\pi}{\sqrt{3}}\right)$. $\bm{\Gamma}$ is the maximum point, the valleys $\pm \bs{K}$ corresponds to minima and $\bm{M}_i$ are saddle points. By tracking the behavior of the Hessian at these points, we arrive at the diagram displayed in Fig. \ref{fig:intro}-(b) (only the region $\phi = [0, \pi]$ is displayed since the Hessian is invariant under $\phi \rightarrow 2\pi - \phi$). Analysis of critical points reveals:
\\
\noindent
\textit{Regions I, I' and I''}: In these, $\bs{M}_{i}$ are saddle points. In I, $\Gamma$ is a maximum and $\pm \bs{K}$ are minima; this pattern reverses in I' and I''. The boundary between regions I and II is defined by
$
t = \tau_{12}(\phi)=\frac{1}{32\sin^2\phi}\left[\cos\phi + \sqrt{\cos^2\phi+32\sin^2\phi}\right],
$
along which the Hessian vanishes, signaling a HOVHS as shown in the first panel of Fig. \ref{fig:intro}-(c) 
In particular, at $\phi = \pi/2 $, 
$\varepsilon_{+}(\bs{M}_1 + \bs{p}) \approx \varepsilon_{+}(\bs{M}_1) + (9/4)p^{2}_{x} - (27/16)p^{2}_{x} p^{2}_{y}$ 
describes a HOVHS with 
$D(\varepsilon) \sim |\varepsilon|^{-1/4}$.
\\
\noindent
\textit{Region II}:
Crossing the boundary into region II, where $t>\tau_{12}(\phi)$, each HOVHS splits into two conventional VHS equidistant from $\bm{M}_{i}$ on the BZ boundary. As we increase $t$ and get closer to the boundary with region III, the VHS tend towards the $\pm \bs{K}$ (second panel of Fig. \ref{fig:intro}-(c)).
\\
\noindent
\textit{Region III:} The boundary between II and III is defined by $t = \tau_{23}(\phi) = 3^{1/4}
{[18\sqrt{3}\sin^2\phi - 18|\sin\phi|\cos\phi}]^{-1/2}$
where the Hessian vanishes when evaluated at the BZ zone corners $\pm \bm{K}$. We note that this curve has an asymptote at $\phi = \pi/6$.
Along the curve $\tau_{23}(\phi)$, $\pm \bm{K}$ are HOVHSs with $D(\varepsilon) \sim |\varepsilon|^{-1/3}$ (third panel of Fig.\ref{fig:intro}-(c)), which separate region II  and III where $\pm \bm{K}$ are local minima or maxima, respectively. As we cross the boundary into region (III), the VHSs split from the $\pm \bm{K}$ points and move towards the center of the BZ (last panel in Fig. \ref{fig:intro}-(c)). 
We note that regions IV and V are similar to III and II, except for the existence of 2 groups of six quasidegenerate VHS, but they are not the focus of this work.
\\

\begin{figure}
    \centering
    \includegraphics[width=8.5cm]{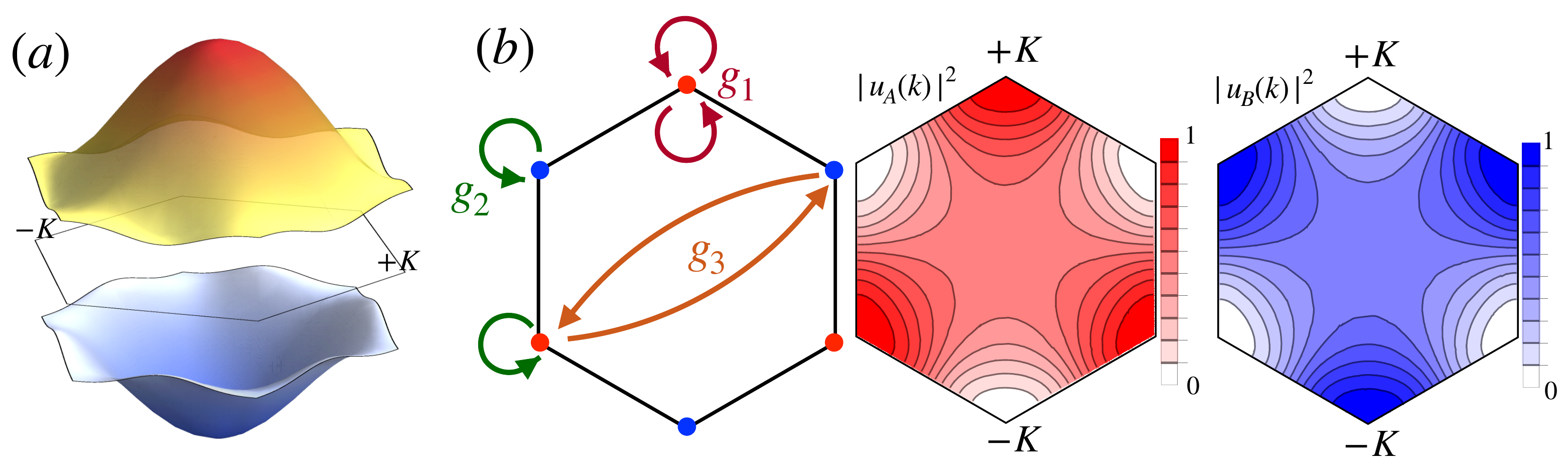}
    \caption{(a) Band structure with HOVHS at $\phi=\pi/2$. (b) Two patch model of the HOVHS and the sublattice weight for the upper band. 
    }
    \label{fig:phasediagram}
\end{figure}

\noindent
\textit{Two HOVHS patches:}
Henceforth, we focus on the boundary between II and III, where  a pair of HOVHS are located at $\pm \bs{K}$ for each band. A typical band structure for $\phi=\pi/2$ is presented in Fig.\ref{fig:phasediagram}(a). Without loss of generality, we focus on the upper band (orange band in the figure). Close to HOVHS filling, we can effectively expand the dispersion within small patches centered at $\pm\bm{K}$ up to fourth order:
\begin{equation}
\begin{split}
&\,
\varepsilon_{\pm\bm{K}}(\bm{p}) = \varepsilon_0 \pm\kappa_1(p_y^3-3p_x^2p_y)-\kappa_2(p_x^2+p_y^2)^2 
\,,
\end{split}\label{eq:epsilon}    
\end{equation}
where $\varepsilon_0$ is the energy at the HOVHS,  
$ \kappa_1 = 
\frac{9t\cos\phi}{8}+\frac{\sqrt{3}|\sin\phi|(6t^2+\csc^2\phi)}{16t}$ and 
$\kappa_2 = 
\frac{27}{64}t\cos\phi - \frac{|\sin\phi|(162t^4+27t^2\csc^2\phi-\csc^4\phi)}{128\sqrt{3}t^3}
$
are two
coefficients that vary continuously along the boundary between II and III, determined by the curve $t = \tau_{23}(\phi)$. 
$\kappa_2 \neq 0$ quantifies the deviation from perfect nesting, which occurs at $(\phi,t) = (\pi/6, \tau_{23}(\pi/6))$. 

This two patch model admits three momentum conserving interactions $g_1, g_2$ and $g_3$ (see Fig.\ref{fig:phasediagram}(b)) defined as
\begin{equation}
     H_I= g_1\psi_\alpha^\dagger\psi^\dagger_\alpha\psi_\alpha\psi_\alpha+\sum_{\alpha\neq\beta}g_2\psi_\alpha^\dagger\psi_\beta^\dagger\psi_\beta\psi_\alpha+g_3\psi_\alpha^\dagger\psi_\beta^\dagger\psi_\alpha\psi_\beta
 \end{equation} 
where we leave the spin structure and spin summation implicit, which should be $\sigma,\sigma',\sigma',\sigma$ for the four fermions operators in each term, and denote the patches by $\alpha,\beta=\pm$ with $\psi_{\pm}=\psi_{\pm{K}+\bm{p}}$. These interactions are obtained from projecting the lattice fermion interactions onto band fermions. Quite interestingly, as long as we focus on patches around $\pm\bm{K}$, the \textit{SI effect always associates each valley with a distinct sublattice index.} We show this effect by plotting the sublattice weights (for the upper band) $|u_{A+}(\bm{k})|^2$ (red) and $|u_{B+}(\bm{k})|^2$ (blue)  in Fig.\ref{fig:phasediagram}(b). 
Crucially, fermions near $+\bm{K}$ ($-\bm{K}$) are solely from A (B) sublattice, so there is a one-to-one correspondence between $c_{i\in A/B}$ and $\psi_{\alpha}$. An important consequence is that the onsite Hubbard interaction $H_U=U\sum_{i} c_{i\uparrow}^\dagger c_{i\downarrow}^\dagger c_{i\downarrow} c_{i\uparrow}$ only contributes  to $g_1$ and the nearest neighbor interaction $H_V=V\sum_{\langle i j \rangle}c^\dagger_{i\sigma}c^\dagger_{j\sigma'}c_{j\sigma'}c_{i\sigma}$ only contributes to $g_2$. {Note that if there were no SI, $H_U$ would contribute to all $g_i$.} Moreover, the SI also implies that \(g_3\) only arises from the more exotic bond-bond interactions \(H_W=W\sum_{\langle i j \rangle}c^\dagger_{i\sigma}c^\dagger_{j\sigma'}c_{i\sigma'}c_{j\sigma}\) \cite{KSSH,KSSH88,NakamuraItoh01}. In the extended-Hubbard model with additional bond-bond interactions ($H_U+H_V+H_W$) the bare interactions before renormalization are therefore
 \begin{equation}
     g_1(0)=U, ~~ g_2(0)=V, ~~ g_3(0)=W.\label{eq:initial}
 \end{equation}
Note that this initial condition is robust against imperfect nesting, as long as the two HOVHS are located at $\pm\bm{K}$. The bond-bond interactions are typically orders of magnitude smaller than other interactions \cite{Hubbard63,GammelCampbell88}, and so we will neglect them below.

\noindent\textit{RG equations:}
To identify the potential electronic instabilities resulting from these interactions in Eq.\eqref{eq:initial}, we note that in the presence of HOVHS the bare particle-hole and particle-particle bubbles at zero momentum or $\pm\bm{K}$ all diverge in the same power-law manner. These bubbles constitute the renormalization of $g_i$'s at one-loop level, resulting in the following RG flow equations \cite{shtyk2017electrons, wu2023pair}:
\begin{equation}
\label{eq:rg}
    \begin{split}
     \dot{g_1}&=\epsilon g_1+(d_2-d_3)g_1^2-2d_2g_2^2\,,\\
     \dot{g_2}&=\epsilon g_2+(d_1-1)g_2^2-2d_2g_1g_2.
     \end{split}
\end{equation}
Here we defined the running parameter as $y=\Pi_{pp}(\bm{q}=0)$, and the nesting parameters are defined as $d_1\approx \Pi_{ph}(\bm{K})/y$, $d_2\approx \Pi_{ph}(0)/y$ and $d_3\approx \Pi_{pp}(\bm{K})/y$. Within our patch model, these bubbles are calculated via 
\begin{equation}
    \begin{aligned}
        &\Pi_{pp}(0)\,, \Pi_{pp}(\bm{K})=\int_{\bm{p}}\frac{1-f[\epsilon_{\bm{K}}(\bm{p})]-f[\epsilon_{\mp\bm{K}}(-\bm{p})]}{\epsilon_{\bm{K}}(\bm{p})+\epsilon_{\mp\bm{K}}(-\bm{p})},\\
        &\Pi_{ph}(\bm{K})=\int_{\bm{p}}\frac{f[\epsilon_{-\bm{K}}(\bm{p})]-f[\epsilon_{\bm{K}}(\bm{p})]}{\epsilon_{\bm{K}}(\bm{p})-\epsilon_{-\bm{K}}(\bm{p})},~\Pi_{ph}(0)=\int_{\bm{p}}\frac{-\partial f[\epsilon]}{\partial \epsilon}
    \end{aligned}\label{eq:bubbles}
\end{equation}
where $\int_{\bm{p}}=\int_{|\bm{p}|<\Lambda}\frac{d^2\bm{p}}{4\pi^2}$ with $\Lambda$ being the high energy cutoff, and in the first line `+' is for $\Pi_{pp}(\bm{K})$ while `-' is for $\Pi_{pp}(0)$.
In the perfect nesting case when $\kappa_2$ in Eq.\eqref{eq:epsilon} vanishes,  
Eq.\eqref{eq:bubbles} can be evaluated analytically, which leads to
$d_1=1$,  $d_2=d_3\approx\frac{1}{3}$. But in general $d_i$'s are determined by both the energy scale and the parameter $\phi$ and can strongly deviate from their perfect nesting values.
The tree-level term \(\epsilon\) results from rescaling the field operators, with \(\epsilon=0\) for the Gaussian fixed point and \(\epsilon=1/3\) for the supermetal fixed point, which occurs at  \(g_2=0\), \(g_1=\frac{1}{3(d_3-d_2)}\). 
Note that the strong coupling fixed points are the same for both choices.

To probe possible symmetry breaking orders, we introduce the following order parameters in both particle-particle and particle-hole channels:
 \begin{equation}
     \begin{aligned}
         &\Delta^s_{\text{SC}}=\langle\psi^\dagger_+\psi^\dagger_-+\psi^\dagger_-\psi^\dagger_+\rangle, ~~ \Delta^t_{\text{SC}}=\langle\psi^\dagger_+\psi^\dagger_--\psi^\dagger_-\psi^\dagger_+\rangle\\
         &\Delta_{\text{PDW}}=\langle \psi_\alpha\psi_\alpha\rangle, ~~\Delta_{\text{CDW}}=\langle\psi^\dagger_+\psi_-\rangle, ~~\Delta_{\text{SDW}}=\langle\psi^\dagger_+\hat{s}\psi_-\rangle\\
         &\Delta_{\text{FM1}}=\sum_\alpha\langle \psi_\alpha^\dagger s_z\psi_\alpha\rangle, ~~\Delta_{\text{FM2}}=\sum_\alpha \alpha\langle \psi_\alpha^\dagger s_z\psi_\alpha\rangle\\
         &\Delta_{\text{PI1}}=\langle \psi_+^\dagger\psi_+ + \psi_-^\dagger \psi_-\rangle, ~~\Delta_{\text{PI2}}=\langle \psi_+^\dagger \psi_+ - \psi_-^\dagger \psi_-\rangle
         \,.
     \end{aligned}
 \end{equation}
 These are singlet and triplet uniform superconductivity $\Delta_{\text{SC}}^{s/t}$, finite momentum superconductivity (pair density wave) $\Delta_{\text{PDW}}$, charge- and spin-density wave $\Delta_{\text{C/SDW}}$, ferromagnetism with different parities $\Delta_{\text{FM1/2}}$ and Pomeranchuk instabilities with different parities $\Delta_{\text{PI1/2}}$. Note that $\Delta_{\text{PI1}}$ amounts to shift the chemical potential by an overall constant, while $\Delta_{\text{PI2}}$ corresponds to shifting the chemical potentials at $\pm\bm{K}$ oppositely. In other words, in the presence of $\Delta_{\text{PI1/2}}$, the system can still remain metallic but is shifted away from the Van Hove filling. When the energy scale is reduced, the order parameters flow as 
 \begin{equation}
    \begin{aligned}
        &\dot{\Delta}_{\text{SC}}^{s/t}=-g_2\Delta_{\text{SC}}^{s/t}, ~~ \dot{\Delta}_{\text{PDW}}=-d_3g_1\Delta_{\text{PDW}},\\
        &\dot{\Delta}_{\text{C/SDW}}=d_1g_2\Delta_{\text{C/SDW}},~~\dot{\Delta}_{\text{FM1/2}}=d_2g_1\Delta_{\text{FM1/2}},\\
        &\dot{\Delta}_{\text{PI1}}=d_2(-g_1-2g_2)\Delta_{\text{PI1}},~~\dot{\Delta}_{\text{PI2}}=d_2(-g_1+2g_2)\Delta_{\text{PI2}}.
    \end{aligned}
 \end{equation}
 Lastly, we will use the renormalized susceptibilities to identify the leading order, which obey $\dot{\chi_i}=d_i|\Delta_i|^2$, where $d_i$ is the corresponding nesting parameters while for the uniform SC we have $d_i=1$. Close to the onset of the instability, $\chi_i\sim (y_c-y)^{\gamma_i}$, thus the most negative $\gamma_i$ corresponds to the leading instability.

\noindent\textit{Phase diagram and two-step RG:}
In Fig.\ref{fig:phasediagramflows} (a) we present the phase diagram for the bare couplings \(g_j(0)\) with tree level contribution for \(\phi=\pi/2\) and \(T=0.001t\). The RG flows are also shown by the stream lines. There are three regions in the phase diagram: two regions with strong coupling instabilities (blue for \(g_2>0\) and red for \(g_2<0\)) in which orders develop, and a region (green) within which no instability of the HOVHS develops and no symmetry is broken, resulting in a supermetal phase. Note that the slopes of the boundaries between the regions are functions of \(\phi\) and \(T\), but all phase diagrams are qualitatively similar. In particular, the boundaries are given by \(g_2=\left(1-d_1\pm\sqrt{(1-d_1)^2+8(3d_2-d_3)}\right)g_1/(4d_2)\)
for \(g_1>0\).

The symmetry broken phases within each region for different values of \(\phi\) and temperature are shown in Fig.\ref{fig:phasediagramflows}(b).  For \(g_2<0\) (region B), 
singlet and triplet superconductivities
are the leading symmetry-breaking instabilities (PI1, corresponding to a chemical potential, is generally the leading vertex correction but does not break any symmetry and does not gap out the Fermi surface). The degeneracy is lifted by the neglected bond-bond interaction giving a non-zero \(W=g_3\), with positive (negative) values favoring 
triplet (singlet)
SC. For \(g_2>0\) (region A), the leading instability is the inversion-symmetry breaking valley-order PI2, with subleading PDW or C/SDW as shown in Fig.\ref{fig:phasediagramflows} (b); observe that C/SDW are Kekul\'{e}-type bond orders due to the sublattice-valley polarization. These are consistent with the phases found in \cite{wu2023pair}.

We note, however, that the PI2 instability does not gap out the HOVHS, but yields an interaction-driven staggered mass in the Haldane model. Once the generated staggered mass \(M\) is sufficiently large, the RG equations Eq. \ref{eq:rg} are no longer valid, as only a single HOVHS at a single valley can be tuned to at a time, resulting in the situation considered in \cite{AksoyChamon23}. In that case the \(g_2\) and \(g_3\) processes become forbidden by momentum conservation. This results in a particularly interesting scenario for the most relevant case of completely repulsive interactions (positive \(g_j\)). The instabilities in RG generally occur at some finite critical value \(y=y_c\), but it is possible that at some point the running \(M(y)\) grows larger than the cutoff \(\Lambda=\Lambda_0 e^{-y}\) (\(\Lambda_0\) being the bare cut-off prior to starting the RG flow) for some \(y^*<y_c\), in which case the RG should be done in two steps as illustrated in Fig.\ref{fig:phasediagramflows}(c). When \(y^*\) is reached, Eq. \ref{eq:rg} has to be modified with \(g_2\rightarrow0\), with the new RG flow continuing along the \(g_2=0\) line in the phase diagram (region C), which is otherwise an unstable fixed trajectory of the RG flow, until \(y_c\) is finally reached.

Depending on whether \(g_1(y^*)\) is positive or negative, this mechanism can realize a Chern supermetal or a broken symmetry phase, respectively. The possible symmetry broken phases are shown in the third panel of Fig.\ref{fig:phasediagramflows}(b), and include either a FM or PDW instabilities. 
We emphasize that these phases would be unstable in the absence of the interaction-generated staggered mass which projects out the \(g_2\) and the previously neglected \(g_3\) interactions. For the supermetal, \(g_3\) in principle leads to a further instability \cite{shtyk2017electrons,wu2023pair} due to multiple HOVHS \cite{classen_competing_2020}.
As the bond-bond interactions \(W=g_3(0)\) are presumed very small, however, the resulting instability may only occur at exceedingly low temperatures 
making the supermetal stable at finite temperature
even in the absence of a generated staggered mass, thanks to the SI mechanism.
The PDW phase is similarly stabilized once \(g_2\) and \(g_3\) are projected out, but importantly \(g_1\) changes sign before the projection, with attraction in the PDW channel thus generated
by purely repulsive bare interactions. This scenario contrasts with models with a single HOVHS \cite{wu2023pair, AksoyChamon23}, where the PDW instability requires bare attractive interactions.

In summary, we have classified the manifold of VHSs in inversion symmetric Haldane bands, uncovering a new framework to investigate electronic correlations in Chern bands. The identification of a pair of HOVHS at BZ valleys provides a route to study the interplay of interactions and band topology once the Fermi energy lies in the vicinity of these higher-order saddles. Beyond a host of exotic ordered phases including long-sought pair-density waves, we identified a novel Chern supermetal displaying non-Fermi liquid behavior and quantum anomalous Hall response, which are salient features observable in transport and tunneling experiments. Expanding this framework to other Chern systems realized either in ultracold fermions or in 2D Van der Waals materials is a promising future direction.
\\

\noindent
\textit{-Note:} During the completion of this work, Ref. \cite{AksoyChamon23} appeared analyzing a single HOVHS in the Haldane model with a staggered mass that explicitly breaks inversion symmetry. Our approach, conversely, starts with an inversion symmetric model with two HOVHSs and leads to different Fermi liquid instabilities.

\section*{Acknowledgments}
We thank to Claudio Chamon, Dmitry Chichinadze and Ajit Srivastava for useful discussions. This research was supported by the U.S. Department of Energy, Office of Science, Basic Energy
Sciences, under Award DE-SC0023327 (D.S. and L.H.S.) and by the Gordon and Betty Moore Foundation’s EPiQS Initiative through GBMF8686 (Y.-M.W).

\bibliographystyle{apsrev4-1}
\bibliography{bibliography}

\end{document}